\documentclass[preprint,12pt,authoryear]{elsarticle}

\usepackage{amssymb}
\usepackage{amsmath}
\usepackage{tcolorbox}
\usepackage{booktabs}

\journal{Nuclear Physics B}

\begin{document}

\begin{frontmatter}



\title{A systematic comparison of Large Language Models for automated assignment assessment in programming education: Exploring the importance of architecture and vendor} 


\author{Marcin Jukiewicz} 
\ead{marcin.jukiewicz@amu.edu.pl}
\affiliation{organization={Department of Logic and Cognitive Science, Adam Mickiewicz University},
            addressline={Wieniawskiego 1}, 
            city={Poznan},
            postcode={61-712}, 
            country={Poland}}

\begin{abstract}
This study presents the first large-scale, side-by-side comparison of contemporary Large Language Models (LLMs) in the automated grading of programming assignments. Drawing on over 6,000 student submissions collected across four years of an introductory programming course, we systematically analysed the distribution of grades, differences in mean scores and variability reflecting stricter or more lenient grading, and the consistency and clustering of grading patterns across models. Eighteen publicly available models were evaluated: Anthropic (claude-3-5-haiku, claude-opus-4-1, claude-sonnet-4); Deepseek (deepseek-chat, deepseek-reasoner); Google (gemini-2.0-flash-lite, gemini-2.0-flash, gemini-2.5-flash-lite, gemini-2.5-flash, gemini-2.5-pro); and OpenAI (gpt-4.1-mini, gpt-4.1-nano, gpt-4.1, gpt-4o-mini, gpt-4o, gpt-5-mini, gpt-5-nano, gpt-5). Statistical tests, correlation and clustering analyses revealed clear, systematic differences between and within vendor families, with “mini” and “nano” variants consistently underperforming their full-scale counterparts. All models displayed high internal agreement, measured by the intraclass correlation coefficient, with the model consensus but only moderate agreement with human teachers’ grades, indicating a persistent gap between automated and human assessment. These findings underscore that the choice of model for educational deployment is not neutral and should be guided by pedagogical goals, transparent reporting of evaluation metrics, and ongoing human oversight to ensure accuracy, fairness and relevance.
\end{abstract}


\begin{highlights}
\item Systematic evaluation of 18 state-of-the-art LLMs reveals distinct grading philosophies: liberal models (GPT-4o, Claude-haiku-3.5) assign higher scores, restrictive models (DeepSeek-Reasoner, GPT-4.1-nano) favor lower grades, and balanced models extensively use intermediate scores
\item Vendor-specific clustering emerges with six distinct evaluation styles: models from the same provider (OpenAI, Google, Anthropic, DeepSeek) demonstrate similar grading patterns, suggesting shared algorithmic approaches to code assessment
\item Limited human-AI agreement challenges automated grading reliability: even the best-performing model (Claude-haiku-3.5) achieves only moderate agreement with human teachers, well below the threshold for good reliability
\item Model size significantly impacts grading quality: full-scale versions consistently outperform their “mini” and “nano” counterparts, with important implications for budget-constrained educational institutions
\item Teachers systematically assign higher grades than all AI models, indicating fundamental differences in evaluation approaches and raising questions about the pedagogical implications of AI-based assessment
\end{highlights}


\end{frontmatter}

\section{Introduction}

The emergence of generative artificial intelligence (GenAI), and in particular large language models (LLMs), has initiated an intensive debate on the future of education. This article draws on and synthesizes the existing research in this area, providing a broad literature overview \cite{202509.1233, Jukiewicz2025}. Since the public release of ChatGPT at the end of 2022, the literature has increasingly depicted a vision of profound and positive—though not unproblematic—transformation, in which intelligent tutoring systems personalize learning paths, while automated assistant tools relieve teachers of time-consuming administrative and didactic tasks \citep{Jukiewicz2025}. One of the most promising application areas is the automation of student work assessment, which makes it possible to deliver immediate, personalized feedback to learners on an unprecedented scale \citep{kiyak2024chatgpt,ruwe2024embracing}.

Research conducted between 2023 and 2025 shows that under specific conditions—especially for short, well-structured tasks with clearly defined criteria—LLMs can achieve accuracy and consistency comparable to human raters \citep{202509.1233,hackl2023,yavuz2025utilizing}. The potential benefits are substantial: automated grading systems can significantly reduce teacher workload, provide students with immediate feedback, and enable scalable assessment in large cohorts.

Despite their tremendous technological potential, LLM-based grading has notable caveats. A broad analysis of 120 scientific articles \citep{Jukiewicz2025} shows that although generative AI offers considerable educational benefits—such as content personalization and increased accessibility—it also raises concerns about academic integrity, the development of critical thinking, and the quality of interpersonal interactions in educational settings. The effectiveness of LLM-based grading depends on task type, model specification, prompting strategy, and educational context. Reports also indicate that generative models perform worse on open-ended programming problems than on tasks with a clearly defined outcome \citep{lee2024can}; some models produce false corrections to code or overlook simple programming errors \citep{azaiz2023ai}, or even “correct” correct solutions \citep{azaiz2023ai}. Repeatability is also problematic—multiple runs of the same task sometimes lead to different scores, suggesting incomplete stability \citep{jukiewicz2024future,gandolfi2025gpt}. In addition, models may hallucinate or generate overly general comments instead of precise feedback \citep{gandolfi2025gpt}. Consequently, most studies emphasize the irreplaceable role of human oversight and human–AI collaboration to ensure accuracy, accountability, and fairness \citep{agostini2024large,yavuz2025utilizing}.

Within rubric-based grading, different LLMs exhibit clearly differentiated effectiveness. Techniques using question-specific rubrics significantly outperform methods based on code similarity (e.g., CodeBERTScore) and question-agnostic rubrics, especially for algorithmically diverse and challenging tasks such as data structures and algorithms; by contrast, in more homogeneous implementations, such as object-oriented programming, both rubric types yield comparable results \citep{pathak2025rubric}. Performance tends to increase with model size, highlighting the benefits of using larger LLMs \citep{zhao2025automated,pathak2025rubric}. Comparative studies show that GPT-4o generally achieves higher consistency and better alignment with human grades than open-source models such as LLaMA 3.2, especially on code-based questions \citep{mendoncca2025evaluating}. Some open-source or smaller models (e.g., DeepSeek, Qwen-PLUS) nevertheless correlate highly with instructors’ grades on C programming tasks, whereas models such as Gemini 2.0 Flash may assign higher or more lenient grades \citep{cisneros2025jorgpt}. Interestingly, off-the-shelf models such as Claude 3.5 Sonnet sometimes outperform models fine-tuned for grading tasks \citep{frias2025flexigrader}. Hybrid approaches that combine LLMs with classical machine-learning algorithms (e.g., Extra Trees Regressor, Random Forest Classifier) can further improve performance, but may still overestimate low grades and underestimate high ones \citep{mahdaoui2025automated}—underscoring the fundamental importance of prompt quality and task definition.

Across fields such as medicine, foreign language learning, computer science, engineering, chemistry and mathematics, LLMs differ not only in their average scores but also in their error profiles. GPT-4 is often more precise but stricter than human raters \citep{grevisse2024llm,thomas2024learning,gandolfi2025gpt}, while GPT-3.5 and some open-source models tend to give higher or more lenient grades \citep{jackaria2024comparative,manning2025,zhuang2024toree}. Some studies highlight a precision–recall trade-off across models: GPT-3.5 achieved high precision but low recall in coding tasks \citep{azaiz2023ai}, whereas GPT-4 reached almost perfect recall but lower precision in error detection \citep{grandel2024}. Several papers report very high agreement indices with human raters \citep{hackl2023,lihuang2024,cohn2024,yavuz2025utilizing}, but agreement varies strongly across models and tasks: tuned or fine-tuned versions consistently outperform “base” models \citep{jin2024using,menezes2024ai,latif2024}.

For several reasons, programming education represents a particularly compelling domain for studying LLM-based assessment. First, programming assignments usually have more objective correctness criteria than essays, potentially reducing the subjectivity that complicates automated evaluation in other domains. Second, the structure of programming languages provides clear syntactic and semantic frameworks for assessment. Third, programming instruction often involves large student cohorts requiring intensive grading, making automation particularly valuable.

This article contributes to the growing body of research on the use of Large Language Models in automated assessment of programming assignments, and more broadly to understanding how different LLMs approach the complex task of educational evaluation, providing insights that extend beyond programming education to other domains where automated evaluation may be helpful. It constitutes a systematic examination of the grading behaviours of different models. It covers the actual distribution of awarded points (0, 0.5, and 1), quantitative comparisons of mean scores and their variability to show which models tend to grade more strictly or leniently. In addition, it assesses the consistency and relationships of grading patterns across models. The analysis also identifies characteristic grading styles and groups of models exhibiting convergent evaluation tendencies (e.g., more restrictive versus more balanced grading), thereby providing an empirical basis for comparing and interpreting the scores generated by LLMs. As educational institutions increasingly grapple with questions about integrating AI technologies while maintaining academic quality and equity, systematic evidence on the variability of LLM performance becomes essential for informed decision-making. At the same time, much of the existing empirical evidence concerns models that are now older or potentially outdated—especially in light of the release of systems such as GPT-5 and Claude Opus 4.1. This underscores the importance of continually evaluating emerging models and approaches to ensure that conclusions about automated assessment remain current and relevant.

Building on these considerations, the present study systematically examines differences and similarities in the grading behaviour of Large Language Models. To make this investigation transparent, the key hypotheses were specified a priori as follows:  

\begin{itemize}
    \item \textbf{H1:} LLMs differ in the distribution of grades (0, 0.5, 1).  \\
    Previous studies have shown that different models vary in grading strictness and tend to prefer different grade levels (e.g., awarding “1” more often than “0.5”), which stems from differences in architectures, training instructions, and default settings \citep{202509.1233,grevisse2024llm,gandolfi2025gpt}.

    \item \textbf{H2:} LLMs differ in mean scores and variability (standard deviation), reflecting differences in grading strictness or leniency.  \\
    Various LLMs exhibit distinct tendencies to grade more strictly or more leniently (model bias literature), hence the expected differences in means and dispersions \citep{grevisse2024llm,gandolfi2025gpt}.

    \item \textbf{H3:} Grading patterns of some LLMs are positively associated.  \\
    Models with similar architectures (e.g., the same vendor or comparable size) may produce similar grades, as suggested by prior comparisons \citep{202509.1233,zhuang2024toree,gandolfi2025gpt,sobo2025evaluating}.

    \item \textbf{H4:} Agreement between some LLMs exceeds chance level.  \\
    For well-defined tasks, even different models tend to converge on similar grades; their agreement should thus be higher than random \citep{202509.1233,mendoncca2025evaluating,sobo2025evaluating,hackl2023,lihuang2024,yavuz2025utilizing}.

    \item \textbf{H5:} Median grades across all 18 models are not equal.  \\
    Since distributions and means differ, the medians should also differ—especially for tasks of varying difficulty \citep{zhuang2024toree,grevisse2024llm,gandolfi2025gpt}.

    \item \textbf{H6:} Specific pairs of LLMs differ significantly in their grading patterns.  \\
    Pairwise comparisons of models with different architectures or grading strategies should reveal significant differences \citep{zhuang2024toree,grevisse2024llm,gandolfi2025gpt,sobo2025evaluating}.

    \item \textbf{H7:} LLMs can be grouped into meaningful clusters based on grading similarity, indicating convergent evaluation styles (e.g., stricter vs.\ more balanced).  \\
    Clustering based on grading correlations has already been applied to both humans and algorithms; models with similar grading styles are expected to form clusters \citep{202509.1233,Jukiewicz2025}.
\end{itemize}

\section{Materials and Methods}

This section outlines the methodological framework of the study. The data were collected from programming tests conducted in an introductory programming course for cognitive science students. Data collection spanned a period of four years. Each year, students completed nine tests covering similar topics and types of tasks. In total, approximately 6500 task records (questions together with student responses) were gathered. After excluding submissions with no answers provided, 6081 complete question–answer pairs were retained for analysis. A detailed description of the course structure and its implementation has been presented in an earlier article \citep{jukiewicz2024future}. The topics and programming concepts addressed during the classes are summarized in Table~\ref{tab2}.

\begin{table}[!h]
\caption{List of topics discussed during the classes \citep{jukiewicz2024future}.}\label{tab2}%
\begin{tabular}{p{0.3\textwidth}p{0.65\textwidth}}
\toprule
\textbf{Lesson} & \textbf{Topic} \\
\midrule
If-else instructions & variables, if, elif, else, comparison operators, logical operators, built-in functions \\
Loops & for, while, break, pass, continue, nested loops, built-in functions \\
Random and time modules & random and time modules \\
Lists, tuples and dictionaries & lists, tuples, dictionaries: creation, indexing, modification, iteration, applications \\
String module & string module \\
Functions: basics & functions: definition, invocation or call, parameters, local and global variables \\
Functions: recursion & recursion \\
Object-oriented programming & class, object, attributes, methods, encapsulation, polymorphism, inheritance \\
Graphical User Interface & tkinter: creating the main window, widgets, interface layout, event handling, graphics and pictures \\
\bottomrule
\end{tabular}
\end{table}

The performance of 18 Large Language Models from four major vendors was compared. These represented the most recent publicly available releases as of August 2025. The models included lightweight variants (e.g., mini, nano, lite) and full-scale flagship versions (e.g., opus, pro, gpt-5), thus covering a representative spectrum of contemporary architectures. The complete list is as follows:

\begin{itemize}
    \item \textbf{Anthropic:} claude-3-5-haiku (20241022), claude-opus-4-1 (20250805), claude-sonnet-4 (20250514).
    \item \textbf{Deepseek:} deepseek-chat, deepseek-reasoner.
    \item \textbf{Google:} gemini-2.0-flash-lite, gemini-2.0-flash, gemini-2.5-flash-lite, gemini-2.5-flash, gemini-2.5-pro.
    \item \textbf{OpenAI:} gpt-4.1-mini, gpt-4.1-nano, gpt-4.1, gpt-4o-mini, gpt-4o, gpt-5-mini, gpt-5-nano, gpt-5.
\end{itemize}

In general, this type of evaluation relies on two prompting strategies. In zero-shot prompting, the model receives only the question, without additional contextual instructions, to probe its raw knowledge. In Chain-of-Thought (CoT) prompting, the model is explicitly instructed to reason step by step before producing the final answer.

In the present study, the second approach was consistently applied. The prompt in Box~1 operationalizes this protocol by requiring the model to work out its own solution and then conduct a structured comparison with the student's submission before assigning a grade.

The grading prompt was prepared based on the author's earlier experiences \citep{jukiewicz2024future, Jukiewicz2025} and subsequently adapted to the tasks analyzed in this study. The full content of the prompt is presented in Box~1.

\begin{tcolorbox}[title=Box 1: The content of the used prompt,label=listing,colback=gray!5,colframe=black!75,width=\textwidth]
\label{box:prompt}
Your task is to grade the student's solution. 

Steps:\\
1. Work out your own solution to the problem. Don't include it in the output.\\
2. Compare your solution to the student's solution. \\
3. Decide if the student's solution is correct. \\
4. Return ONLY a valid JSON object with the following structure:\\

\{\\
  ''is same as actual'': ''yes'' or ''no'',\\
  ''student grade'': ''correct'' or ''almost correct'' or ''incorrect'',\\
  ''feedback'': ''Plain English feedback here, no Markdown, no code fences.''
\}\\

Question: \{question\}\\
Student's solution: \{student\_solution\}\\
Actual solution: steps to work out the solution and your solution here

\textbf{Curriculum scope constraint (very important):}\\
Evaluate the student only on concepts taught up to and including module \{n\} for this project. Do not require or assume knowledge from later classes. Suppose an improvement would involve a later concept. In that case, you may mention it as an optional advanced suggestion, but do not deduct points.

\textbf{Course sequence (in order):}\\
1. Conditional statements
2. Loops
3. Pseudorandom numbers \& time library
4. Array-like types (lists, tuples, sets, dictionaries)
5. Strings
6. Functions
7. Functions
8. Object-oriented programming
9. GUI

\textbf{Assumed knowledge for this project (modules 1--\{n\}):}\\
\{assumed\}

\textbf{Important:}\\
- Deductions and justifications must reference only the assumed knowledge.\\
- Lack of later concepts must not reduce the grade.
\end{tcolorbox}

A key feature of the prompt is that the model must first generate its solution to the given problem and then compare it with the student's submission. The student's work is classified as correct, almost correct, or incorrect, and assigned 1, 0.5, or 0 points, respectively. Detailed descriptions of the three possible grades are provided in Table~\ref{tab:grades}.

\begin{table}[!h]
\caption{An extended description of each of the three grades that the student could receive \citep{jukiewicz2024future}.}
\label{tab:grades}
\begin{tabular}{p{0.2\textwidth}p{0.7\textwidth}}
\toprule
Grade & Grade description\\ \midrule
Correct & The student fully understood the given task and met all specified requirements. 
The program runs as expected and fulfills all test conditions.\\
Almost correct & The student understood some aspects of the task, but certain parts may be unclear or omitted. 
The code contains shortcomings that affect the program's functionality. 
The program may work correctly in some cases but fails in others.\\
Incorrect & The student did not grasp the essence of the task or its key aspects. 
The code includes serious errors that prevent the program from functioning correctly.\\
\bottomrule
\end{tabular}
\end{table}

\section{Results}
Building on the methodological framework described above, this section presents the results of our study of Large Language Models' grading behaviour. It explains how the distributions of scores assigned by the models were analysed, how differences in grading tendencies and consistency were quantified, and how distinctive evaluation styles and clusters of models exhibiting similar patterns were identified. The chapter outlines these aspects and provides a roadmap for detailed findings. It sets the stage for a nuanced comparison of model-based assessment with human evaluation.

\subsection{Descriptive statistics}

\begin{figure}[!h]
    \centering
    \includegraphics[width=1\textwidth]{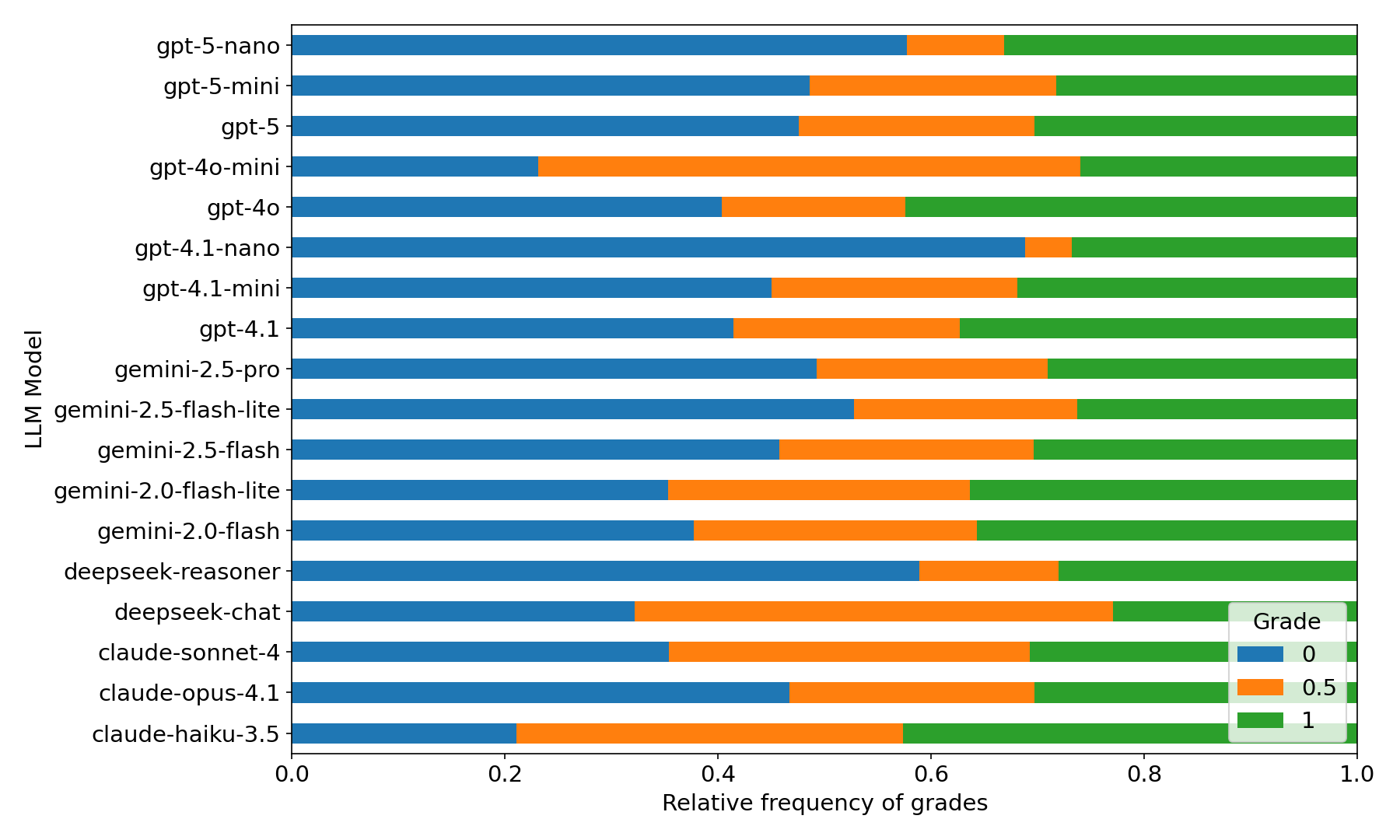}
    \caption{Relative frequency of grades assigned by different LLMs.}
    \label{fig:distribution_barplot}
\end{figure}

As illustrated in Figure~\ref{fig:distribution_barplot}, the models diverge in their overall strictness and use of intermediate grades. These proportions were computed as the number of responses of a given grade (0, 0.5 or 1) divided by the total number of analysed tasks for each model.  

Models such as gpt-4o, its mini variant, and claude-haiku-3.5 exhibited a relatively high proportion of maximum scores (1.0), suggesting a more liberal grading approach and a tendency to recognize solutions as correct more frequently. By contrast, models such as deepseek-reasoner and gpt-4.1-nano were dominated by failing grades (0.0). This reflects a more restrictive grading strategy that often classifies student responses as incorrect. Between these two extremes lies a group of models that extensively used the intermediate category (0.5). For example, claude-sonnet-4 and gemini-2.0-flash-lite more often acknowledged partial correctness, indicating a balanced grading style with greater sensitivity to intermediate levels of accuracy.  

These findings have direct practical implications, as the choice of model for automated grading may substantially influence student outcomes. The diversity of grading profiles indicates no single consistent evaluation standard; instead, the models exhibit distinct, systematic patterns in their grading behavior.  


Table~\ref{tab:model_summary} complements the distributional analysis by providing absolute counts of grades (0, 0.5, and 1) and the mean score and standard deviation for each model. While the scores constitute ordinal data, we report mean values as an additional descriptive measure to illustrate overall grading tendencies. In situations where the median alone does not differentiate between distributions, the mean can reveal whether the prevailing tendency is more lenient or stricter (e.g., median 0.5 with mean 0.7 vs. median 0.5 with mean 0.2).
  
These descriptive statistics, therefore, add nuance to the previously reported proportions by showing how often models used each grade in absolute terms and how consistent they were in their grading behavior.

\begin{table}[!h]
\centering
\caption{Summary of grading results per model (counts, mean, and standard deviation).}
\label{tab:model_summary}
\begin{tabular}{lrrrrr}
\toprule
Model & Count 0 & Count 0.5 & Count 1 & Mean & Std. Dev. \\
\midrule
claude-haiku-3.5   & 1281 & 2207 & 2593 & 0.608 & 0.384 \\
claude-opus-4.1    & 2838 & 1399 & 1844 & 0.418 & 0.431 \\
claude-sonnet-4    & 2151 & 2060 & 1870 & 0.477 & 0.406 \\
deepseek-chat      & 1956 & 2729 & 1396 & 0.454 & 0.368 \\
deepseek-reasoner  & 3579 &  798 & 1704 & 0.346 & 0.440 \\
gemini-2.0-flash   & 2293 & 1617 & 2171 & 0.490 & 0.428 \\
gemini-2.0-flash-lite & 2145 & 1723 & 2213 & 0.506 & 0.423 \\
gemini-2.5-flash   & 2783 & 1452 & 1846 & 0.423 & 0.429 \\
gemini-2.5-flash-lite & 3206 & 1276 & 1599 & 0.368 & 0.424 \\
gemini-2.5-pro     & 2995 & 1316 & 1770 & 0.399 & 0.431 \\
gpt-4.1            & 2518 & 1295 & 2268 & 0.479 & 0.443 \\
gpt-4.1-mini       & 2739 & 1401 & 1941 & 0.434 & 0.434 \\
gpt-4.1-nano       & 4186 &  264 & 1631 & 0.290 & 0.442 \\
gpt-4o             & 2451 & 1051 & 2579 & 0.511 & 0.455 \\
gpt-4o-mini        & 1404 & 3096 & 1581 & 0.515 & 0.350 \\
gpt-5              & 2893 & 1345 & 1843 & 0.414 & 0.433 \\
gpt-5-mini         & 2955 & 1409 & 1717 & 0.398 & 0.426 \\
gpt-5-nano         & 3508 &  557 & 2016 & 0.377 & 0.461 \\
\bottomrule
\end{tabular}
\end{table}

\subsection{Correlation analyses}

\begin{figure}[!h]
    \centering
    \includegraphics[width=\textwidth]{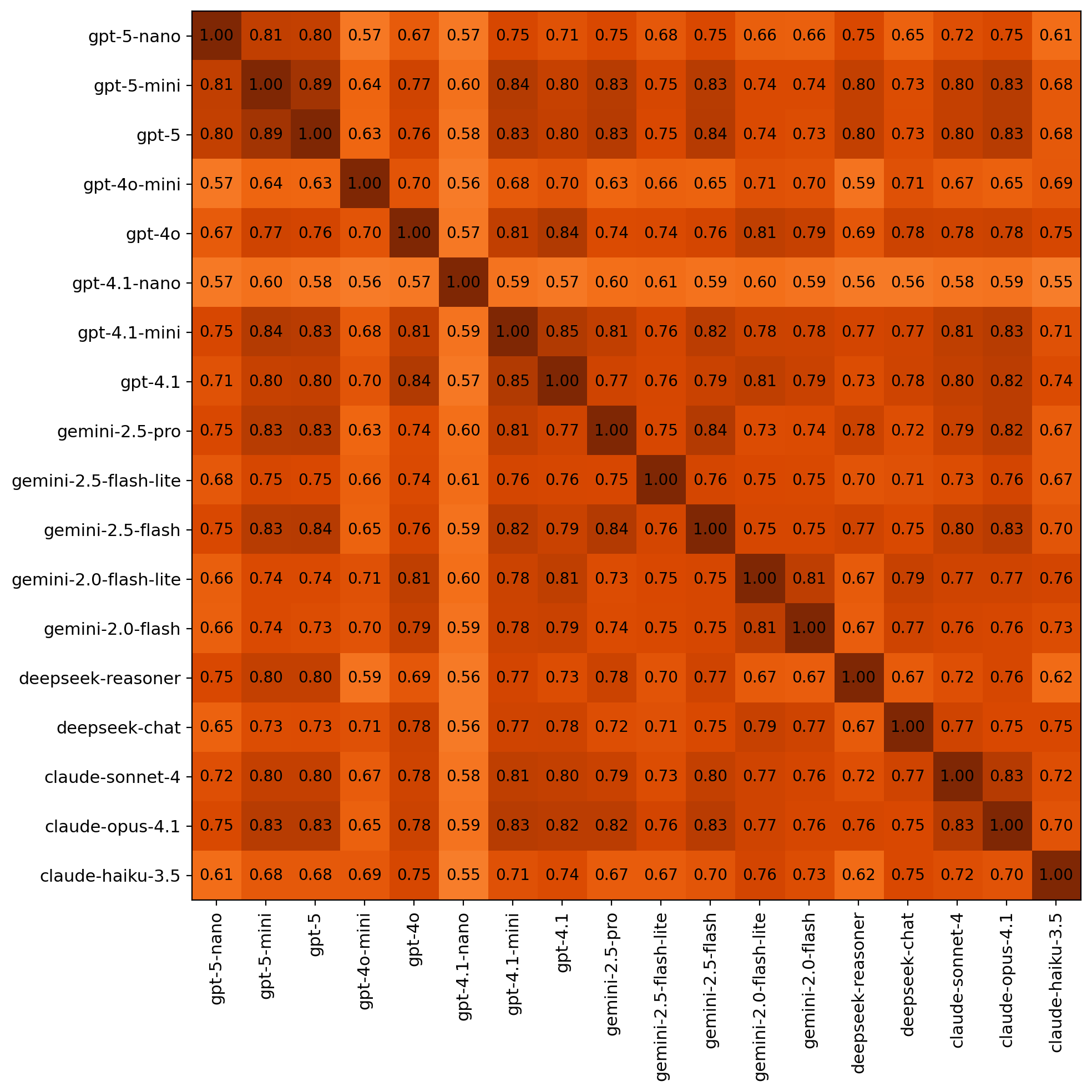}
    \caption{Spearman correlation heatmap (clipped to 0–1) for all LLM models.}
    \label{fig:spearman_heatmap}
\end{figure}

The Spearman rank correlation analysis (Figure~\ref{fig:spearman_heatmap}) reveals that all models exhibit positive associations in their grading behavior, with coefficients ranging from approximately 0.55 to 0.89. The highest correlations are observed within groups of models provided by the same vendor, particularly between gpt-5-mini, gpt-5, and gpt-5-nano ($\rho \approx 0.80$--$0.89$), as well as within the Gemini family (e.g., gemini-2.5-flash, gemini-2.0-flash, and gemini-2.5-pro, $\rho \approx 0.79$--$0.83$). Strong associations are also found among models from the gpt-4.1 series (e.g., gpt-4.1 vs gpt-4.1-mini, $\rho \approx 0.85$).

Lower correlation values are observed for models that deviate more strongly in grading style, such as claude-haiku-3.5 and gpt-4.1-nano, which reach coefficients in the $0.55$--$0.68$ range compared to other systems. These results suggest that while all models exhibit some degree of convergent grading, substantial differences exist between individual model families.

\begin{figure}[!h]
    \centering
    \includegraphics[width=\textwidth]{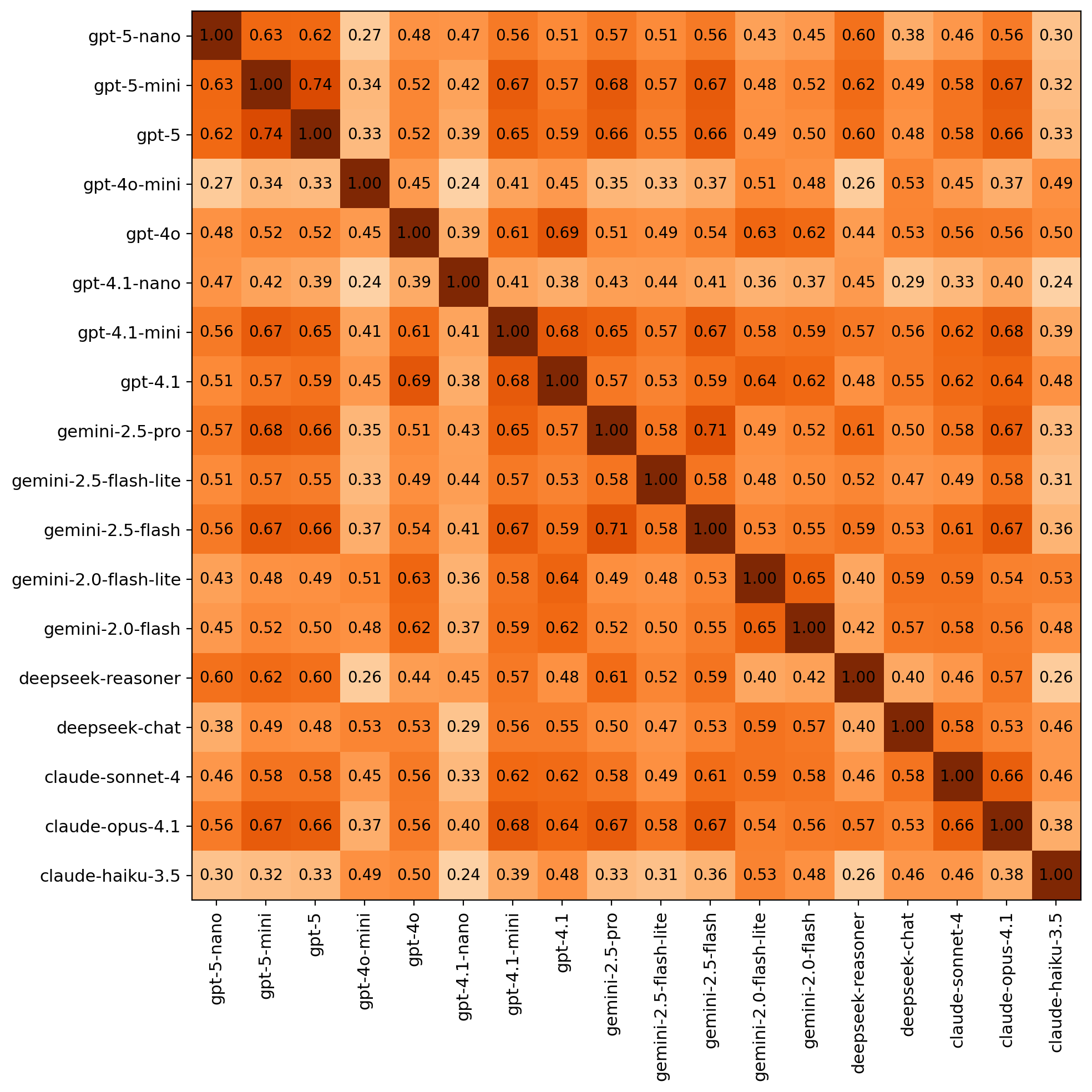}
    \caption{Cohen’s $\kappa$ agreement heatmap between LLMs (values clipped to $[0,1]$).}
    \label{fig:cohen_kappa_heatmap}
\end{figure}

The Cohen 's~$\kappa$ agreement matrix (Figure~\ref{fig:cohen_kappa_heatmap}) reveals a more heterogeneous picture of model similarities than the Spearman correlation. $\kappa$ values range widely, from very low ($\approx 0.20$--$0.30$) to moderate and high ($\approx 0.60$--$0.74$). The strongest agreement is again found within families of models from the same vendor, particularly among the gpt-5 group (e.g., gpt-5 vs.\ gpt-5-mini, $\kappa \approx 0.74$) and between gpt-4.1 and gpt-4.1-mini ($\kappa \approx 0.68$). At the same time, $\kappa$ values are consistently lower than the corresponding Spearman coefficients. This highlights that while grading patterns follow similar trends, point-by-point agreement on exact scores is more limited.

The lowest $\kappa$ values are observed in comparisons involving models such as claude-haiku-3.5 and gpt-4.1-nano, which deviate substantially from the rest. Their $\kappa$ rarely exceeds~0.30, indicating poor grading agreement despite these models still showing positive rank correlations.

Spearman and Cohen 's~$\kappa$ analyses demonstrate a discrepancy between similarity in ''grading style'' and actual agreement in assigned grades. Models within the same family (e.g., gpt-5, gpt-4.1, or Gemini) display high correlations and relatively high $\kappa$ values, reflecting consistent evaluation. By contrast, cross-vendor comparisons, particularly involving claude-haiku-3.5, show positive correlations (models move in the same direction) but low $\kappa$ values, indicating substantial divergence in final score assignments.

These findings suggest that while LLMs converge in their overall grading tendencies, they often fail to achieve high consistency at the level of individual assessments. 

\subsection{Statistical comparisons}

The Friedman test revealed significant differences among the performance scores of the evaluated language models ($\chi^2_{(17)} = 10630.24$, $p < 0.001$). Thus, the models' performance levels were statistically significantly different.

The post hoc analysis (Conover test with Holm and Holm-Sidak corrections) showed that out of all pairwise model comparisons (18 models, totaling 153 pairs), only 7 pairs did not exhibit a statistically significant difference in evaluation scores ($p \geq 0.05$ after correction). Conversely, for the vast majority of comparisons (approximately 95\% of pairs), significant differences were observed ($p < 0.05$ after correction). The model pairs for which no significant differences were found are listed in Table~\ref{tab:nosig-en}. Applying Holm and Holm-Sidak adjustments yielded consistent significance outcomes – the same set of pairs remained non-significant regardless of the correction method.

\begin{table}[htbp]
\centering
\caption{Pairs of language models for which no significant differences in evaluation scores were found (Conover post hoc test with Holm and Holm-Sidak correction, $p \geq 0.05$).}
\label{tab:nosig-en}
\begin{tabular}{ll}
\hline
Model A & Model B \\
\hline
gemini-2.5-flash & claude-opus-4.1 \\
gemini-2.5-flash & gpt-5 \\
gpt-4.1 & claude-sonnet-4 \\
gpt-4.1 & gemini-2.0-flash \\
gpt-5-mini & gemini-2.5-pro \\
gemini-2.0-flash-lite & gpt-4o \\
claude-opus-4.1 & gpt-5 \\
\hline
\end{tabular}
\end{table}

Notably, the above model pairs have comparable evaluation quality – the lack of significant differences suggests that the models in each pair achieve similar performance. For example, the models gemini-2.5-flash, claude-opus-4.1, and gpt-5 form a group with statistically indistinguishable performance (no pairwise comparison among them reached significance). Likewise, no significant differences were observed between gpt-4.1 and claude-sonnet-4, or between gpt-4.1 and gemini-2.0-flash; similarly, the pairs of gpt-5-mini with gemini-2.5-pro and gemini-2.0-flash-lite with gpt-4o showed no significant differences. These results indicate that aside from the noted exceptions, most models differ significantly in quality, enabling a clear ranking of model performance. At the same time, the presence of a few model pairs with no significant differences suggests the existence of subsets of models with closely comparable performance levels, which may have implications for model selection in specific applications or for further analysis of their characteristics.

These results strengthen the interpretation of clustering and correlation analyses, showing that vendor-specific LLM groups correlate more strongly and differ less frequently in post-hoc contrasts. At the same time, cross-vendor comparisons highlight substantial divergences in grading strictness and style.

\subsection{Clustering analysis}

Clustering analysis of the language models' grading results was conducted using the $k$-means method and hierarchical clustering. 

\begin{figure}[!h]
    \centering
    \includegraphics[width=0.85\linewidth]{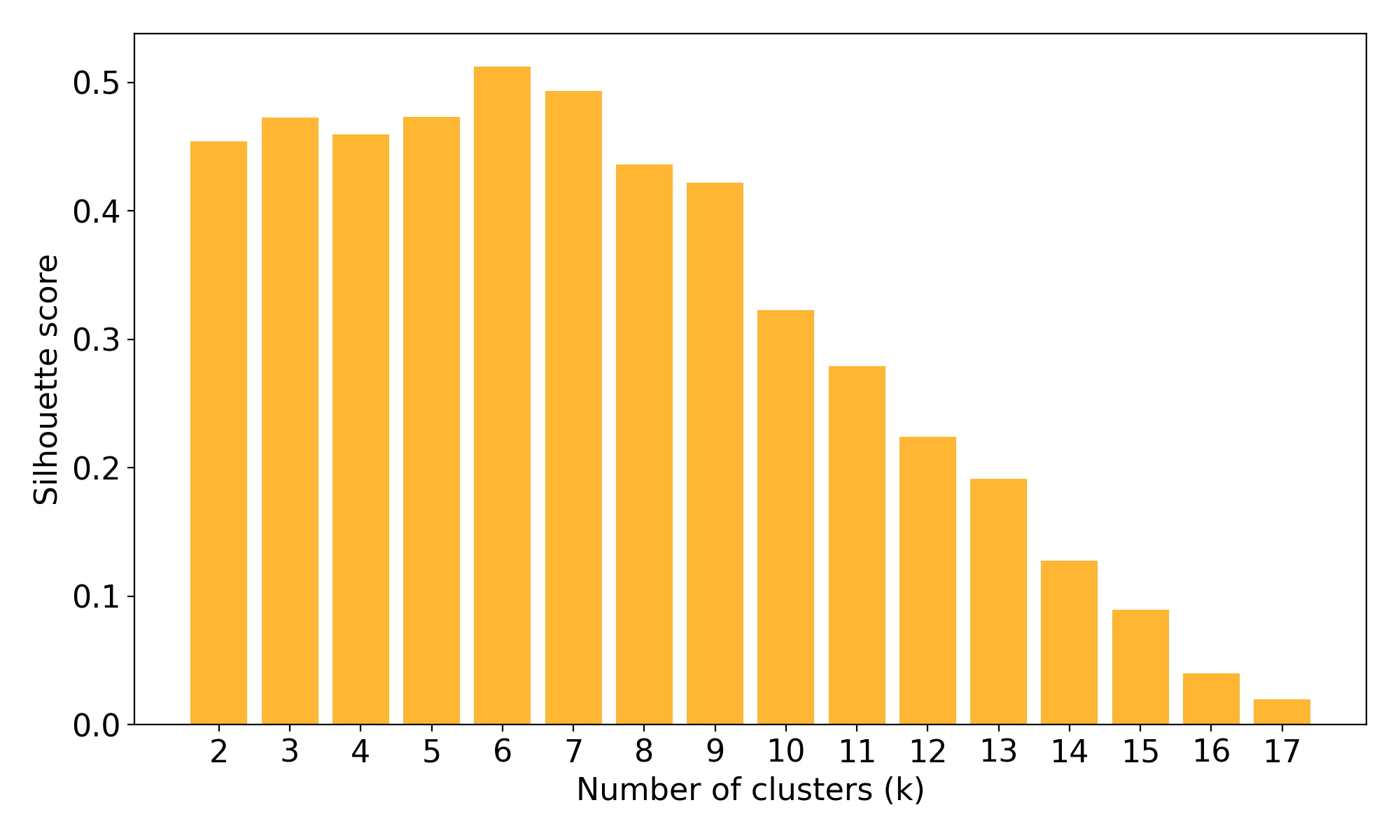}
    \caption{Silhouette score for different numbers of clusters ($k$) in the $k$-means analysis of grade distributions.}
    \label{fig:silhouette}
\end{figure}

In the first step, the silhouette score plot for cluster counts $k$ from 2 to 17 indicated an optimal number of clusters of 6 (Figure~\ref{fig:silhouette}). The silhouette coefficient, measuring cohesion and separation, reached its maximum at $k=6$ (approximately $0.5$), suggesting a transparent partition of the data into six well-separated groups.

\begin{figure}[!h]
    \centering
    \includegraphics[width=0.95\linewidth]{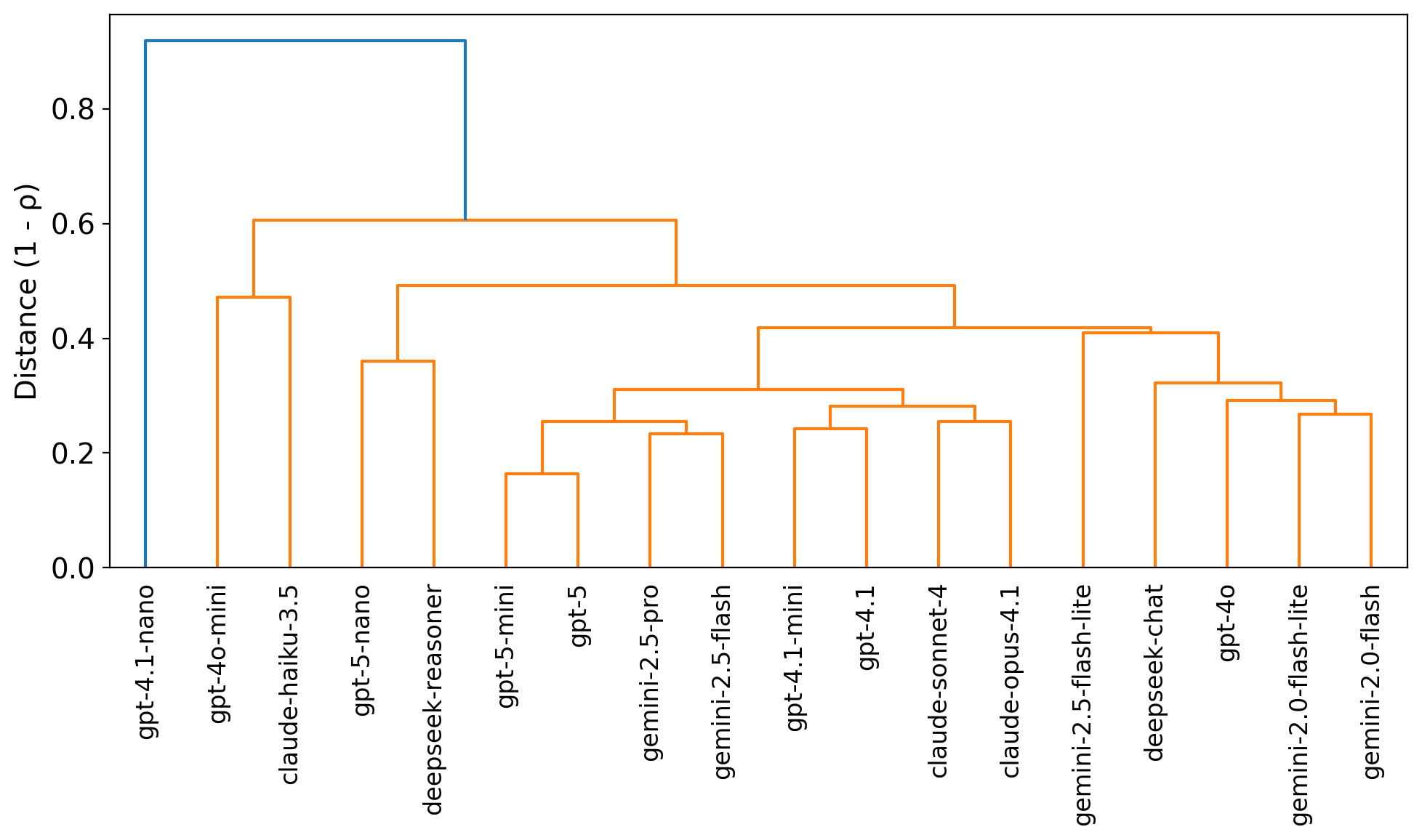}
    \caption{Hierarchical clustering dendrogram based on Spearman correlations between models' grading patterns.}
    \label{fig:dendrogram}
\end{figure}

Next, hierarchical clustering was performed based on the models' similar scoring behavior. The distance metric used was $1 - \rho$, where $\rho$ is the Spearman rank correlation coefficient computed between the scoring patterns of each pair of models. The dendrogram (Figure~\ref{fig:dendrogram}) shows that models group into clusters corresponding to similar scoring styles. Distinct clusters of models with high mutual score correlation can be identified, indicating that these models assign identical ratings to the same items.

The identified clusters include:
\begin{itemize}
    \item \textbf{gpt-5 cluster:} models gpt-5, gpt-5-mini, and gpt-5-nano, which exhibit similar behavior (primarily giving high scores of $1$ with very few $0$ scores).
    \item \textbf{gpt-4 cluster:} models gpt-4.0, gpt-4.0-mini, gpt-4.1, and gpt-4.1-mini, characterized by a more moderate scoring style (a balanced distribution of $0.5$ and $1$ scores).
    \item \textbf{Gemini cluster:} the Gemini family of models (including gemini-2.5-pro, gemini-2.5-flash, gemini-2.5-flash-lite, gemini-2.0-flash, gemini-2.0-flash-lite), which tend to assign predominantly high scores of $1$ (i.e., they are more ''lenient'' in their evaluations).
    \item \textbf{DeepSeek cluster:} models deepseek-reasoner and deepseek-chat, distinguished by frequent use of the intermediate score $0.5$ (a relatively cautious or balanced evaluation style).
    \item \textbf{Claude cluster:} models claude-sonnet-4 and claude-opus-4.1, which—similarly to the gpt-5 and Gemini clusters—more often give scores of $1$ than $0.5$ or $0$.
    \item \textbf{Outlier cluster:} models gpt-4.1-nano and \textbf{claude-haiku-3.5}, which do not fit into the above groups. gpt-4.1-nano issues the score $0$ far more frequently (a much stricter evaluator), whereas claude-haiku-3.5 predominantly uses the $0.5$ score (avoiding giving the highest score $1$). These two models are grouped separately due to their atypical scoring profiles.
\end{itemize}

\begin{figure}[!h]
    \centering
    \includegraphics[width=0.95\linewidth]{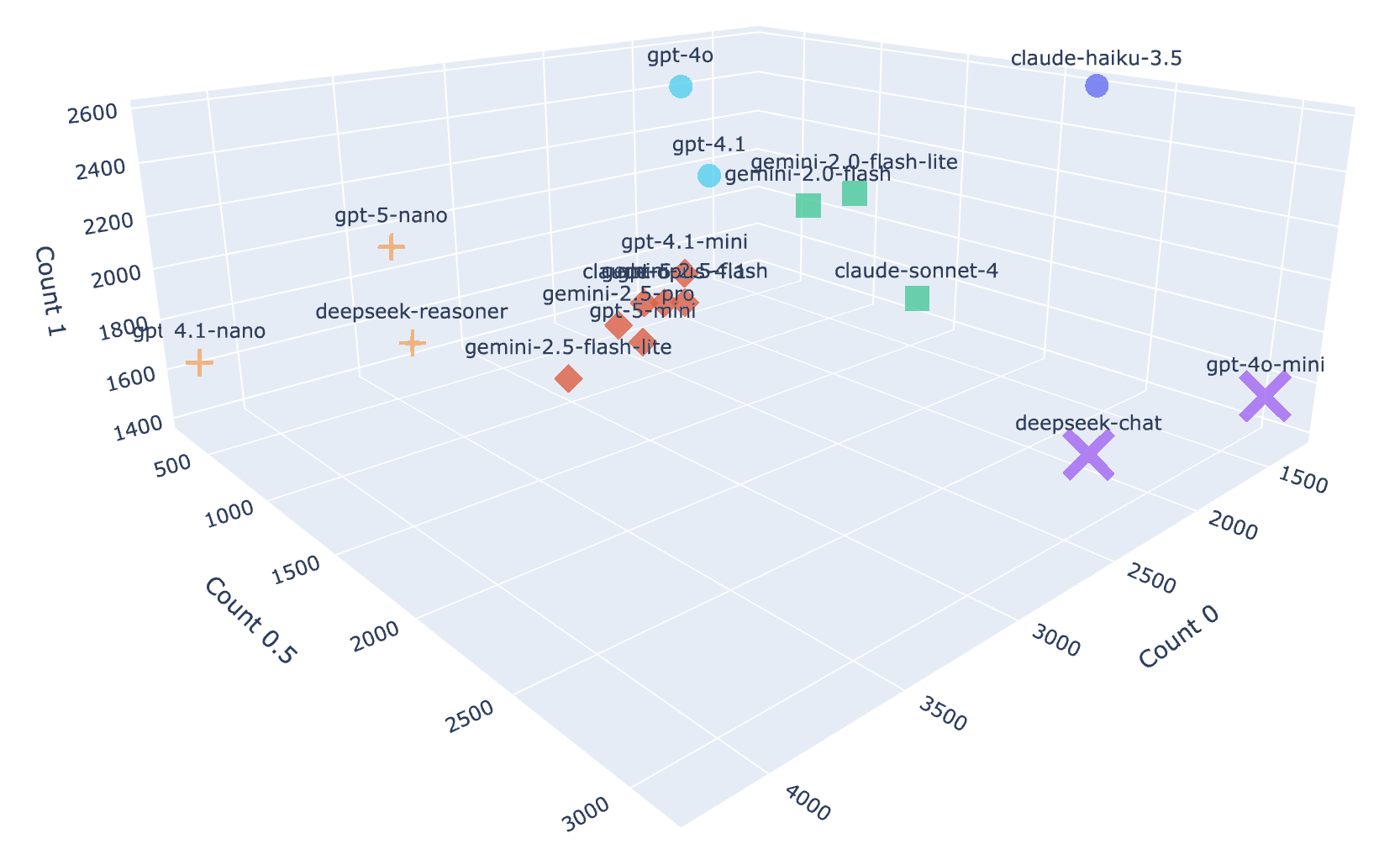}
    \caption{Three-dimensional visualization of model clusters based on counts of $0$, $0.5$, and $1$ scores. Each point represents a model, positioned according to its grade distribution. Colors and marker shapes indicate cluster membership.}
    \label{fig:3d_clusters}
\end{figure}

The cluster assignments obtained via the $k$-means method (for $k=6$) are consistent with the above structure—the $k$-means cluster centroids correspond to similar groupings of models as observed in the hierarchical dendrogram. In the three-dimensional visualization of the score distribution (Figure~\ref{fig:3d_clusters}), one can see that models belonging to the same cluster lie close to each other and their cluster's centroid. These centroids reflect the characteristic score profile of each cluster: for example, the gpt-5 and Gemini cluster has a centroid shifted toward the $1$ axis (indicating a high proportion of score $1$), the DeepSeek cluster's centroid has its most significant component along the $0.5$ axis (frequent intermediate scores), and the outlier cluster (containing gpt-4.1-nano together with claude-haiku-3.5) has a centroid indicating a relatively high number of unusual or extreme ratings (a dominance of $0$ or $0.5$ instead of $1$, respectively).

These six clusters thus reveal apparent differences in the evaluation styles of the various model groups. Models within the same cluster demonstrate high agreement in their scoring (strong correlation or statistical agreement, as evidenced by Spearman correlations and Cohen's kappa measures within the group). In contrast, there are more pronounced discrepancies in rating the same items between clusters. Identifying these clusters has important implications for further analysis of LLMs' agreement and scoring styles: it allows understanding which models evaluate similarly (and thus could be considered redundant sources of judgment) versus which models represent different approaches to evaluation. This, in turn, enables future studies to focus on comparisons across clusters (e.g., to detect systematic differences in evaluation criteria) and improve the consistency of scoring within and between clusters.

\subsection{Comparison of model grades with human grades}

Teachers obtained the highest mean score (0.726, SD = 0.391) and the highest median score (1, occurring 3 854 times, 63.38\% of all cases) among all grading sources. Figure 1 shows the corresponding distributions for the automated models, which can be compared with the teachers’ results reported here. This indicates that teachers were more inclined to award higher scores than the language models. None of the models reached a mean close to this level, suggesting a systematic difference between human and automated assessments. In practice, this means that the models tended to be stricter, assigning lower scores (0 or 0.5) more frequently, while teachers rewarded students with higher grades more often. This observation provides an important reference point for interpreting the analyses presented in this chapter, as it highlights that even when models show strong internal consistency—expressed here by high intraclass correlation coefficient (ICC) values among them—their collective consensus still diverges from the teacher-established standard.

The highest level of agreement was obtained by claude-haiku-3.5, with an ICC value of 0.470, corresponding to a moderate level of agreement. This is the highest score among the compared models. However, it still does not indicate substantial agreement (i.e., it does not reach the threshold for the substantial category \citep{shrout1979intraclass,koo2016guideline}). 

The lowest agreement was observed for gpt-4.1-nano (ICC = 0.204), which, according to standard criteria, corresponds to only slight agreement, close to the lower boundary of the slight category. Such a low result suggests that this model's responses were almost random relative to the expected reference. 

Most other models achieved results in approximately 0.30–0.43, which can be classified as fair (moderately low) to the lower boundary of moderate agreement. For example, gemini-2.0-flash (0.433) and gpt-4o (0.438) reached around 0.43 (moderate agreement), while deepseek-reasoner (0.295) and gemini-2.5-flash-lite (0.311) obtained results closer to 0.3 (weaker, fair agreement). 

Differences between models suggest that larger or more advanced versions produce more consistent grades than their simplified counterparts. For instance, the full gpt-4.1 (0.421) significantly outperforms gpt-4.1-nano (0.204). Similarly, gpt-5 (0.378) performs better than gpt-5-nano (0.314), and gpt-4o (0.438) outperforms gpt-4o-mini (0.365). The ''mini'' and ''nano'' versions generally achieve lower agreement values, suggesting that reducing model complexity diminishes consistency. 

Within the Claude family, the best performer was claude-haiku-3.5 (0.470), surpassing even the newer variants claude-sonnet-4 (0.430) and claude-opus-4.1 (0.382). This suggests that in this particular metric, haiku 3.5 was exceptionally well-tuned or better suited to the task. 

It is worth noting that according to standard ICC criteria, all obtained values (below 0.5) formally indicate poor reliability of measurement \citep{shrout1979intraclass,koo2016guideline}. In other words, none of the models reached the threshold considered good agreement (e.g., ICC $\geq$ 0.75 is regarded as ''good'' agreement). This shows that while some models outperform others, the overall consistency of their responses with the reference grades is not high. Detailed results are presented in Table~\ref{tab:icc_results_en}.

\begin{table}[!h]
\centering
\caption{Comparison of model grades with human grades (ICC(2,1)).}
\label{tab:icc_results_en}
\begin{tabular}{lc}
\toprule
Model & ICC(2,1) \\
\midrule
claude-haiku-3.5   & 0.470 \\
claude-opus-4.1    & 0.382 \\
claude-sonnet-4    & 0.430 \\
deepseek-chat      & 0.395 \\
deepseek-reasoner  & 0.295 \\
gemini-2.0-flash   & 0.433 \\
gemini-2.0-flash-lite & 0.428 \\
gemini-2.5-flash   & 0.394 \\
gemini-2.5-flash-lite & 0.311 \\
gemini-2.5-pro     & 0.366 \\
gpt-4.1            & 0.421 \\
gpt-4.1-mini       & 0.392 \\
gpt-4.1-nano       & 0.204 \\
gpt-4o             & 0.438 \\
gpt-4o-mini        & 0.365 \\
gpt-5              & 0.378 \\
gpt-5-mini         & 0.361 \\
gpt-5-nano         & 0.314 \\
\bottomrule
\end{tabular}
\end{table}

\subsection{Agreement with model consensus as reference}

Table~\ref{tab:icc_consensus} presents each model's intraclass correlation coefficient ICC(2,1) values when the reference grade was defined as the most frequent grade across the 18 models. The highest agreement with this consensus grade was obtained by gpt-4.1-mini (ICC = 0.892), followed closely by claude-opus-4.1 (0.888) and gpt-5-mini (0.877). Most models reached ICC values above 0.8, indicating good agreement with the consensus. The lowest scores were observed for gpt-4.1-nano (0.599) and claude-haiku-3.5 (0.664), reflecting only moderate alignment with the rest of the models. Adopting the aggregated model grade as the reference results in higher agreement levels than those observed with human teachers' grades, suggesting that discrepancies in the traditional approach may have stemmed from errors or inconsistencies in human grading.

This type of analysis raises methodological concerns. It assumes that models themselves define the correct grade, whereas teacher assessments are usually considered the gold standard. Suppose the majority of models share the same systematic error. In that case, their agreement does not guarantee correctness—it may only reflect consistent repetition of the same mistake. Prior research stresses that when disagreements are considerable, aggregation (e.g., majority vote) does not necessarily improve reliability, and assuming a single ''truth'' established by the majority may be misleading \citep{wong2022ground}. Evaluating models against their own consensus risks circular reasoning: models similar to the group appear more accurate, while outliers are penalized, even if sometimes correct.

On the other hand, consensus among models can be a reasonable reference if teacher grades are error-prone. As reported in previous studies, teachers may be fatigued, inattentive, or overly lenient \citep{jukiewicz2024future}. Suppose the original ''gold standard'' is unreliable. In that case, the collective judgment of many independent models may offer more stable and reproducible results. In this sense, 18 models act as a ''wisdom of the crowd,'' where aggregation reduces random individual errors. As shown in Table~\ref{tab:icc_consensus}, consensus-based reference values increased ICC for leading models, suggesting that models are more consistent with each other than with human graders. This approach also highlights outliers such as gpt-4.1-nano and claude-haiku-3.5, which diverged from the majority. While not a replacement for a human gold standard, this additional analysis sheds light on the internal consistency of LLMs and possible weaknesses in teacher-based reference grading.

\begin{table}[!h]
\centering
\caption{ICC(2,1) values for each model with the model consensus (majority vote) as the reference grade.}
\label{tab:icc_consensus}
\begin{tabular}{lc}
\toprule
Model & ICC(2,1) \\
\midrule
claude-haiku-3.5   & 0.664 \\
claude-opus-4.1    & 0.888 \\
claude-sonnet-4    & 0.836 \\
deepseek-chat      & 0.781 \\
deepseek-reasoner  & 0.791 \\
gemini-2.0-flash   & 0.811 \\
gemini-2.0-flash-lite & 0.808 \\
gemini-2.5-flash   & 0.874 \\
gemini-2.5-flash-lite & 0.803 \\
gemini-2.5-pro     & 0.860 \\
gpt-4.1            & 0.864 \\
gpt-4.1-mini       & 0.892 \\
gpt-4.1-nano       & 0.599 \\
gpt-4o             & 0.822 \\
gpt-4o-mini        & 0.672 \\
gpt-5              & 0.871 \\
gpt-5-mini         & 0.877 \\
gpt-5-nano         & 0.782 \\
\bottomrule
\end{tabular}
\end{table}

\section{Discussion}

The obtained results confirm hypotheses H1 and H2, indicating significant differences in the distribution of grades and the mean scores across models. This heterogeneity has profound practical implications for implementing automated grading in educational environments. Particularly noteworthy are the differences between ''liberal'' models (such as gpt-4o and claude-haiku-3.5), which more frequently assigned maximum grades, and ''restrictive'' models (such as deepseek-reasoner and gpt-4.1-nano), which dominated in assigning negative grades. This polarization suggests that models may have different built-in tendencies to recognize solution correctness or tolerance thresholds for incomplete answers. The presence of ''balanced'' models, which extensively used the intermediate category (0.5 points), indicates a third distinct approach to grading. Models such as claude-sonnet-4 and gemini-2.0-flash-lite demonstrated greater sensitivity to partial correctness. This may be particularly valuable in educational contexts where acknowledging student progress is as important as identifying errors.

Similar patterns can be observed in the existing literature, which likewise classifies Gemini 2.0 and OpenAI o1 as ''harsh'' raters and notes ChatGPT 3.5's greater leniency \citep{jiaocomparing}, as well as neutral or slightly lenient behaviour in Claude 3.5 and Gemini 1.5 Pro depending on the task \citep{jiaocomparing}. Comparable but somewhat narrower effects have also been described in \citep{cisneros2025jorgpt}.

Correlation and clustering analyses revealed strong grouping patterns consistent with model families provided by the same vendor, confirming hypotheses H3, H4, and H7. The highest Spearman correlations were observed within the gpt-5 and Gemini families, suggesting that models from the same vendor share similar algorithmic approaches to grading. Identifying six optimal clusters provides a practical framework for understanding the landscape of automated grading. Each cluster represents a distinct evaluation philosophy:

\begin{itemize}
    \item \textbf{gpt-5 cluster:} characterized by a strong tendency to assign maximum grades,
    \item \textbf{gpt-4 cluster:} demonstrating a more balanced approach, with a distribution of medium and high scores,
    \item \textbf{Gemini cluster:} showing a more lenient grading style, dominated by high scores,
    \item \textbf{DeepSeek cluster:} favoring a cautious approach with frequent use of intermediate scores,
    \item \textbf{Claude cluster:} (except for the outlier) also tending toward higher grades,
    \item \textbf{Outlier cluster:} consisting of models with atypical grading profiles.
\end{itemize}

These results are consistent with previous comparative studies showing systematic differences within and between vendor families. Earlier evaluations demonstrated that even different versions of models from the same provider can yield divergent outcomes \citep{sobo2025evaluating}. That models from various vendors may exhibit distinct grading tendencies reflecting their internal mechanisms \citep{grevisse2024llm}. Conversely, reports of high correlations between gpt-4o, gpt-4.1, and Gemini \citep{cisneros2025jorgpt} further overlap with the strong intra-family correlations and clustering patterns observed in this study, which may support the view that models from the same provider tend to share similar approaches to grading and can be meaningfully grouped into distinct evaluation styles. However, because comparable sets of models have rarely been examined side by side, it isn't easy to relate these findings unequivocally to earlier research.

Hypotheses H5 and H6 addressed fundamental statistical differences among language models in the context of automated grading of programming tasks. Confirmation of both hypotheses has essential implications for understanding the nature of automated grading in educational contexts. Demonstrating systematic differences between models globally (H5) and in pairwise comparisons (H6) indicates that the choice of a particular language model for grading tasks is not technically neutral but carries pedagogical consequences that may significantly affect student outcomes.

One of the most notable findings is the limited agreement between all evaluated models and the human teachers' grades. The highest ICC value (0.470 for claude-haiku-3.5) still falls within the category of ''moderate'' agreement, well below the threshold typically considered ''good'' reliability (ICC $\geq 0.75$). This outcome has several possible interpretations. First, it may point to fundamental differences in how humans and machines evaluate programming code. Teachers may consider context, student intent, partial understanding of concepts, and pedagogical factors not directly accessible to language models. The teachers' mean grade was the highest among all compared sources, suggesting a systematic difference in grading leniency. Second, this result may reflect limitations in the design of the prompt or in how models interpret grading criteria. Although the prompt was designed based on the author's prior experience, it is possible that not all nuances of pedagogical grading were effectively communicated to the models.

An interesting perspective emerges from the analysis that uses model consensus as the reference point. When the most frequent grade among the 18 models was taken as the standard, most models achieved high ICC values ($>0.8$), suggesting strong internal consistency among models. This paradox—high agreement among models but low agreement with humans—requires cautious interpretation. On the one hand, it may indicate systematic errors in teachers' grading due to fatigue, oversight, or leniency, as documented in the literature. In this context, the collective judgment of 18 independent models may provide more stable and reproducible results, operating under the ''wisdom of the crowd.'' On the other hand, high ICC values with model consensus may merely reflect consistent replication of the same systematic errors by similar systems. Suppose most models share the same limitations in understanding code or interpreting grading criteria. In that case, their agreement does not guarantee correctness; it only ensures consistency in error.

The results suggest a relationship between model size or complexity and grading quality. ''Mini'' and ''nano'' versions consistently achieved lower ICC values compared to their full-sized counterparts (e.g., gpt-4.1 vs. gpt-4.1-nano, gpt-5 vs. gpt-5-nano). This pattern has practical implications for implementing automated grading in educational institutions, where budget constraints may encourage using smaller, cheaper models. At the same time, it is noteworthy that claude-haiku-3.5, despite being a lightweight model, achieved the highest agreement with human grades. This suggests that specialized fine-tuning may be more critical than sheer model size, opening opportunities for developing models tailored to educational tasks.

\section{Conclusions}
This study provided the first systematic comparison of Large Language Models in the context of automated grading of programming assignments. The results demonstrate that models differ in grading style, score distributions, and agreement with human teachers. This diversity reflects distinct “grading philosophies” ranging from liberal, balanced, to restrictive approaches. The choice of model should therefore not be arbitrary but aligned with pedagogical goals and the context of instruction.

Statistical analyses, including the Friedman test, Spearman rank correlations, and clustering, revealed consistent grouping patterns according to vendor origin. This indicates that architecture and training methods strongly shape how models evaluate student work. The identified clusters provide a practical framework for interpreting the landscape of automated grading, showing that models from the same vendor often share similar strategies.

The agreement between model-generated grades and human teachers’ assessments proved limited. Even the best-performing systems only reached levels corresponding to moderate reliability, far below thresholds typically considered good. Moreover, the average teacher score was higher than any model's, suggesting systematic differences in grading style. When model consensus was used as a reference, however, very high internal consistency emerged, pointing to stable shared grading schemes while simultaneously raising the possibility of replicating the same systematic shortcomings.

A clear relationship was also observed between model size and grading quality, with full-scale versions showing stronger consistency than their “mini” or “nano” counterparts. This has practical implications for educational institutions where the choice of model is often constrained by resources. Notably, claude-haiku-3.5, despite being lightweight, achieved the highest agreement with teachers, highlighting the importance of specialization and fine-tuning for educational tasks.

This study delivers a large-scale, side-by-side comparison of contemporary Large Language Models in the automated grading of programming assignments. The analyses revealed apparent, systematic differences in grading style, strictness and leniency, and the level of agreement with human teachers. Vendor-specific clustering showed that models from the same family tend to share evaluation strategies. At the same time, “mini” and “nano” versions consistently underperformed their full-scale counterparts, suggesting that both architecture and model size shape grading quality. At the same time, even the best systems achieved only moderate agreement with teachers’ scores, highlighting a persistent gap between human and automated assessment. High intraclass correlation coefficients among models signal stable internal schemes and raise the risk of replicating shared biases. These findings underscore that the choice of model for educational deployment is not neutral: institutions should align model selection with pedagogical goals, maintain human oversight, and regularly re-evaluate emerging systems to ensure accuracy, fairness, and relevance as the technology evolves.

\section*{Declarations}

\subsection*{Acknowledgements}

I would like to thank P.L. for invaluable remarks that improved this manuscript.

\subsection*{Funding}

This work was supported by the Polish National Science Centre (grant no. 2025/09/X/HS6/00038). 

\subsection*{Conflict of interest}

I declare that I have no competing interests relevant to the content of this article.




\bibliographystyle{elsarticle-harv}







\end{document}